\documentclass[reprint,superscriptaddress,nobibnotes,amsmath,amssymb,aps,prm,floatfix]{revtex4-1}
\usepackage{multirow}
\usepackage{graphicx}
\usepackage{dcolumn}
\usepackage{bm}
\usepackage{xcolor}
\usepackage{textcomp}
\usepackage{hyperref}
\hypersetup{colorlinks=true, linkcolor=blue, filecolor=blue, urlcolor=blue, citecolor=blue }
\newcommand{\etal}[1]{\textcolor{black}{{\textit{et al.}}}}
\begin{document}

\title{Atomic Structure and Modifiers Clustering in Silicate glasses: Effect of Modifier Cations}

\author{Achraf Atila}
\email{achraf.atila@fau.de}
\affiliation{Friedrich-Alexander-Universit\"at Erlangen-N\"urnberg, Materials Science and Engineering, Institute I, Martensstr. 5, Erlangen 91058, Germany}

%\date{\today}

\begin{abstract}
Oxide glasses are made of a network of glass former polyhedra, and modifiers which have a role in neutralizing the charge of the glass former polyhedra or depolymerize the glass network. The effect of the modifier content on the structure and properties of the glass are to some extent, well known. However, the effect of the type of modifiers on the clustering and the tendency to form a phase separation in the glass is not investigated in detail and still not fully understood until now. Here, we use molecular dynamics to investigate the effect of the modifier type on the structure and the clustering tendency of a series of modified silicate glasses. Specifically, we show that the tendency of modifier---modifier cluster formation is linked to the modifier size and modifier---oxygen bond strength. The effect of different modifiers on the short- and medium-range structure of the glass is also discussed. This allows us to get an overview of the effect of cations nature on the properties of the glass and opens a new window for further development and optimization of the glass properties.

\end{abstract}

%\keywords{Silicates glasses, Field strength, Modifiers clustering, Phase separation, Structural properties.}
\maketitle

\section{\label{sec:Introduction}INTRODUCTION}
Silicates glasses are present in our daily life in many forms, such as window glasses \cite{Luo2016, Wang2019}, protective cover glasses \cite{Yu2018}, glass fibers \cite{Mauro2013, Luo2016}, and laboratory glassware \cite{Smedskjaer2014}. The structure of silica glass is formed by a continuous network of SiO$_4$ tetrahedra connected by bridging oxygen (BO). The addition of network modifiers results in depolymerization of the glass network, by breaking the Si---O---Si bonds and transforming the BO to non-bridging oxygen (NBO) linked to only one silicon and free oxygen (FO) and sometimes to oxygen triclusters (TBO), as in the case of alkaline and alkaline-earth aluminosilicates. This depolymerization affects remarkably the properties of the glass \cite{Atila2019a, Ghardi2019, Bauchy2012, Smedskjaer2010b,Smedskjaer2010a, Kjeldsen2013, Lu2020}.

The development of new oxide glasses with improved properties is affected by the lack of knowledge of the atomic structure of the glass \cite{Atila2019b, Smedskjaer2011}. Numerous studies on silicate glasses modified by different types of cations (e.g., alkali and/ or alkaline earth, rare-earth metals) were performed in order to establish some structural models to describe the glass structure or to find a composition-structure-properties relationship \cite{Lusvardi2007, Bauchy2014a, Xiang2013, Pnitzsch2016}. Moreover, the interpretation and even the prediction of the glass properties  depend strongly on the structural models proposed in the literature to describe the glass structure and the distribution of modifiers in it \cite{Maass1999, Greaves2007, Dyre2009}. 

The existence of some local heterogeneity within the glass matrix and network of modifiers is still unclear. Also, there is some debate about the spatial distribution of the network modifiers; are they homogeneously distributed throughout the glass network, or they show some clustering and a heterogeneous distribution in the form of micro-segregation like the modifier channels in sodium silicate glasses \cite{Greaves1985}. This clustering of modifiers is crucial as it remarkably affects the mobility of the modifiers and plays a vital role in phase separation \cite{Greaves1995, Greaves2007, Bauchy2011}. Moreover, knowing that silicate glasses exhibit phase separation, which is mainly attributed to the clustering of the modifiers \cite{Du2003, Mead2006, Tilocca2007, Christie2010, Christie2012, Jardnlvarez2018, Tour2019, Wang2019, Atila2019a}. It is of great importance to investigate the effect of different modifiers on the structure, know how different modifiers are distributed in the glass matrix, and reveal the driving force behind such behavior. 

The cation field strength (FS) as defined by Dietzel \cite{Dietzel1942} FS = Z$_{X}$/(r$_{X}+$r$_{O}$)$^2$, where Z$_X$ and r$_X$ stand for the charge and ionic radius of the cations, respectively, indicates indirectly the cation---oxygen bond strength (higher FS have higher bond strength). This parameter could be used to study the effect of different modifiers on the structure and properties of oxide glasses. The oxygen ionic radius is omitted from the FS equation, as it is approximated to be the same (FS = $\frac{Z_X}{r_X^2}$) \cite{Weigel2016, Atila2020a, Atila2019b}.

In order get some insights into the atomic structure of the glass, we use molecular dynamics (MD) simulations, which is an effective method to help to investigate and understand the atomic-scale behavior of materials and have been applied in studying and oxide glasses \cite{Atila2020a, Atila2019a, Ghardi2019, liu2018, Mantisi2015}.  We study the effect of the modifier type on the structural properties of a series of modified silicate glasses were the concentration of the modifiers was fixed to 25 mol\%. We compare the change in the structure induced by different modifiers into the glass network and also trends in modifiers clustering with FS. 

The remainder of this paper is organized as follows. In Sec. \ref{sec:Methods}, we give a description of the simulation procedure, the interatomic potential used in our simulations, and we provide a brief description of the different tools used to analyze the structure. In Sec. \ref{sec:Results}, we present the results were we show the effect of the modifier type on the structure and modifiers clustering in silicate glasses. Sec. \ref{sec:Discussion} contains a discussion of the results, which correlates between structural change and modifier cations field strength. Concluding remarks are given in Sec. \ref{sec:Conclusion}.

\section{\label{sec:Methods}Computational details and methodology}
We used on the well-established potential developed by Pedone \textit{et al.} \cite{Pedone2006}. This potential gives a realistic agreement with available experimental data, as mentioned in the literature\cite{Atila2020a, Atila2019b, Ghardi2019, Yu2018, Pedone2006, Pedone2008, Turlier2018, Luo2016b}. Potential parameters and partial charges are given in the reference (\cite{Pedone2006}). MD simulations were performed in the large-scale atomic/molecular massively parallel simulator LAMMPS \cite{Plimpton1995}.

We simulated forty silicate glasses (X$_{n/2}^{n+}$O)$_y$-(SiO$_2$)$_{(1 - y)}$, where (y = 0.05, 0.1, 0.15, 0.2, 0.25) and X stands for Li, Na, K, Mg, Ca, Sr, Ba, or Zn, using classical molecular dynamics. All systems consist of approximately 3000 atoms placed randomly in a cubic simulation box with no unrealistic overlap between atoms. The velocity-Verlet algorithm was used to update the velocities and atomic positions with an integration time step of 1 fs. Periodic boundary conditions (PBC) were applied in all directions. The short-range interaction cutoff distance was chosen to be 5.5 \AA\ \cite{Pedone2006} and long-range interactions were evaluated by the Ewald summation method, with a real-space cutoff of 12.0 \AA\, and a precision of 10$^{-5}$.

\begin{figure}[htbt]
\centering
 \includegraphics[width=\columnwidth]{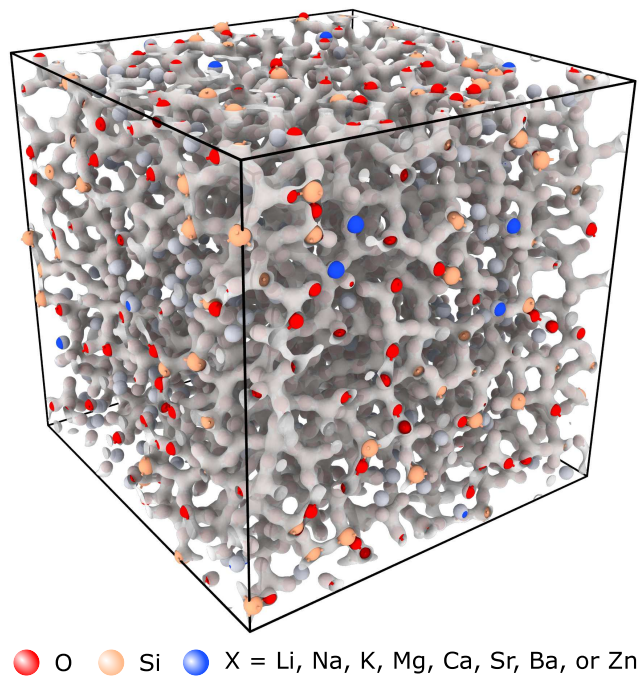}
 \caption{Snapshot of the atomic structure of the studied silicate glasses as visualized by OVITO \cite{Stukowski2009}. The mesh represents the linking between the Si---O.}
 \label{systemvis}
\end{figure}

The systems were melted in NVT ensemble at high temperature (T = 5000 K) for 500 ps, which is enough to bring our systems to the liquid state. The systems were subsequently linearly quenched from T = 5000 K to room temperature (T = 300 K) with a cooling rate of 10$^{12}$ K/s in NVT. Furthermore, the systems were equilibrated at room temperature and zero pressure in NPT for 1 ns make the systems stress free and another supplementary 100 ps in the NVT ensemble for statistical averaging. Nos\'e-Hoover thermostat and barostat were used for temperature and pressure control. Atomic visualization example of the resultant glass is shown in Fig.~\ref{systemvis}.

\section{\label{sec:Results}Results}
\subsection{Local environment of the glass}
\begin{figure}[htbt]
\centering
\includegraphics[width=\columnwidth]{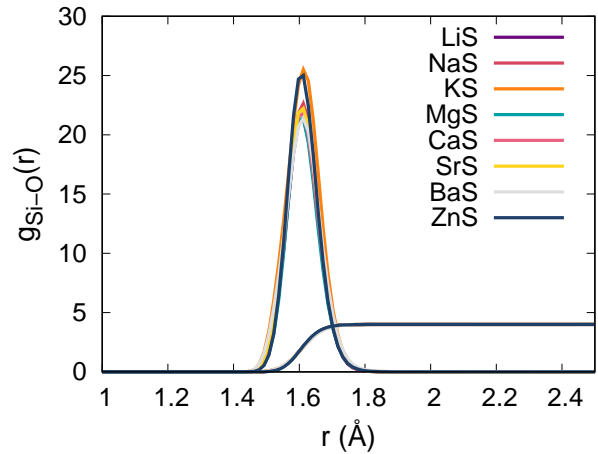}
\caption{RDF and the cumulative coordination function of Si---O pair in the obtained glasses at 300 K.}
\label{fig:SiAlO}
\end{figure}

%In silicate glasses, Si can attain different coordination numbers with oxygen, depending on the thermal and pressure histories of the glass \cite{Stebbins2019a, Stebbins2019b}.
In Fig.~\ref{fig:SiAlO}, we plotted the first peaks of Si---O RDF from the sample with 25 mol\%\, of the modifiers (the Si---O RDF for the other samples are shown in the supplementary materials, Fig. S1). We observe that the first peaks of the Si---O RDF show maxima at a distance around 1.61 \AA\,, which is in agreement with the data reported in the literature \cite{Atila2019a, Atila2019b, Bauchy2014a, Charpentier2018}. Additionally, we see that the first peak position, which represents the mean distance between each pair, is unaffected by the type of the modifier.

By integrating the partial radial distribution functions to a specific cutoff value defined as its first minimum, we can obtain the averaged coordination number Si (number of O around Si). This minimum is found to be 2.1 \AA\, (see Table. \ref{Table:Rij}). The results show that the network former Si have a coordination around 4.0. This is poorly affected by the type of modifier. Moreover, other ab-initio, MD simulations as well as previously reported experiments \cite{Takahiro2018,REN2019} have shown that Si$^{4+}$ is present in tetrahedral coordination (Si$_{4}$), which is in a good agreement with our simulations. 

\begin{figure*}[htbt]
\centering
\includegraphics[width=\textwidth]{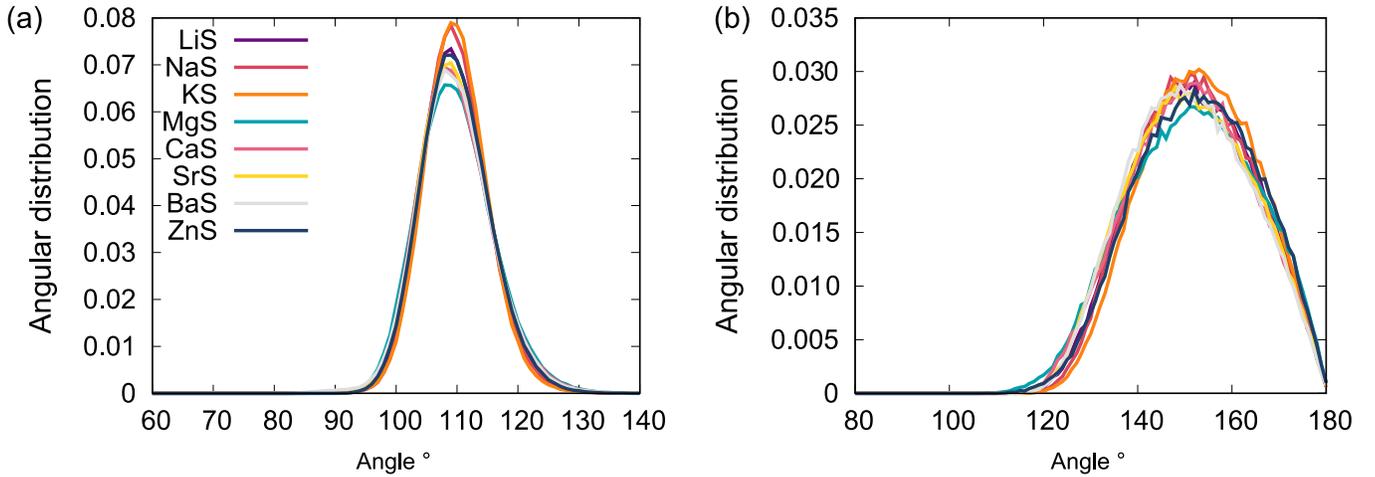}
\caption{Bond angles distributions obtained from MD simulations at 300 K. (a) O-Si-O and (b) Si-O-Si, (c) X-O-X, and (d) X-O-X, where (X = Li, Na, K, Mg, Ca, Sr, Ba, and Zn).}
 \label{SiOSi}
\end{figure*}

The bond angle distributions (BAD) can provide information about inter and intra polyhedral angles (linkage). The O---Si---O BAD presents the distribution of the angles inside SiO$_{4}$ tetrahedra. The bond angle distributions are presented in Fig.~\ref{SiOSi}(a and b). For the O---Si---O distribution, no shift was reported as we changed the modifier and this angle has been found to be centered around a mean value of 108.9\textdegree\, for all systems (Fig.~\ref{SiOSi}(a)). 

The Si---O---Si bond angle distributions presented in Fig.~\ref{SiOSi}(b) provide more insights on the linkage between the SiO$_{4}$ tetrahedra. We observe that Si---O---Si bond angle distribution has a value around 151.6\textdegree\, for all systems independently from the field strength of the modifiers.

\begin{figure}[htbt]
\centering
\includegraphics[width=\columnwidth]{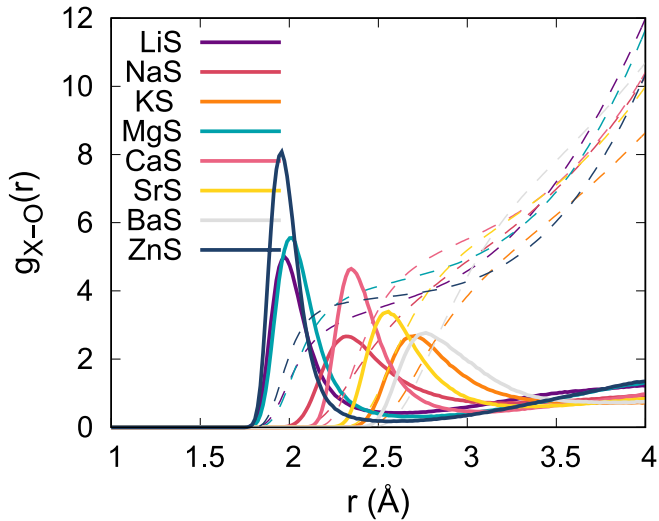}
\caption{X---O, dashed lines are the cumulative coordination numbers of silicate glasses as obtained from MD simulations at 300 K. X = Li, Na, K, Mg, Ca, Sr, Ba, and Zn.}
\label{fig:RDFXO}
\end{figure}

Figure~\ref{fig:RDFXO} shows the RDF function of the modifier---oxygen pairs, the mean separation distance of Li---O, Na---O, K---O, Mg---O, Ca---O, Sr---O, Ba---O, and Zn---O is summarized in Table. \ref{Table:Rij} and it shows that a decrease with increasing field strength. Furthermore, the coordination number decreases with increasing field strength (See Table. \ref{Table:Rij}). %Also, the X---X RDF Fig.~\ref{fig:RDFXO}(b) shows that the X---X distance decreases with increasing FS.

\begin{figure*}[htbt]
\centering
\includegraphics[width=\textwidth]{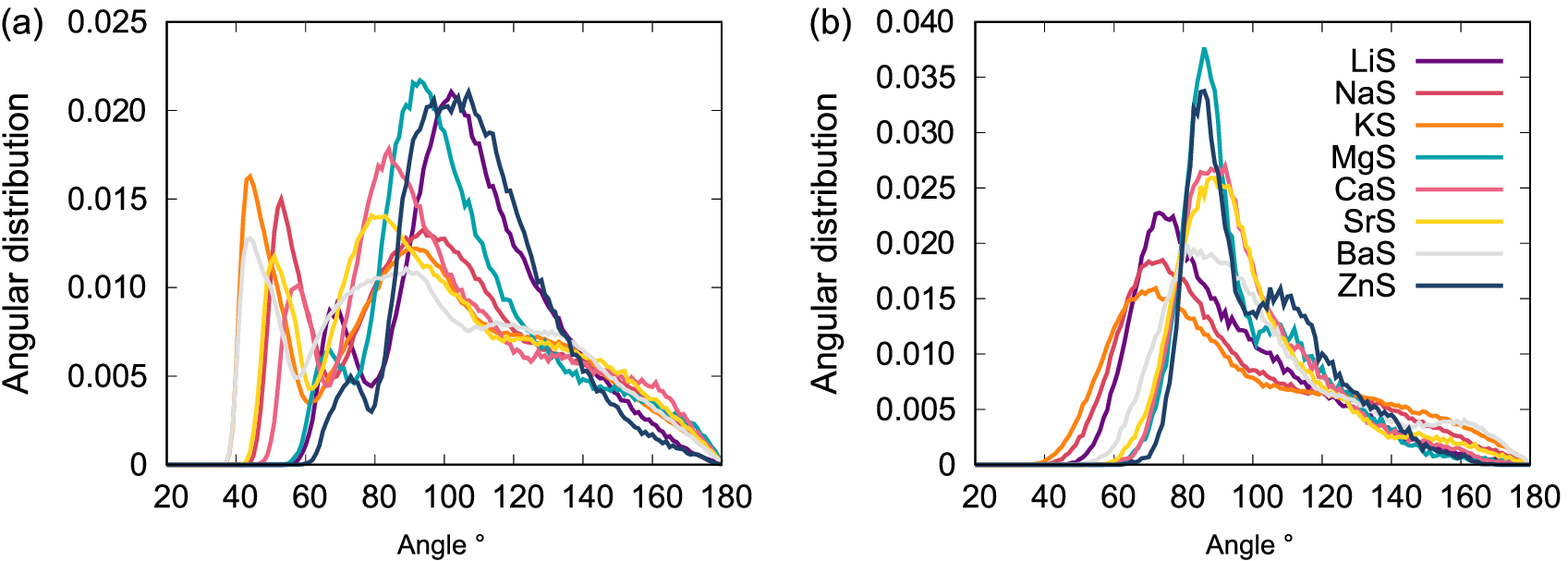}
\caption{Bond angles distributions obtained from MD simulations at 300 K. (a) O---X---O and (b) X-O-X, where (X = Li, Na, K, Mg, Ca, Sr, Ba, and Zn).}
 \label{XOX}
\end{figure*}
Now, if we look at the BAD of O---X---O and X---O---X as plotted in Fig.~\ref{XOX}(a and b), we notice that the environment of the modifiers cation is more complex, which is rather ordinary and expected as the modifiers show a broad distribution in their coordination (the high coordination states that can be attained by these cations which have a maximum of 10 for Ba cations, and a minimum of 3.8 for the Mg cations). For all glasses, the O---X---O BAD show a broad bimodal distribution and as the FS increases, the main peaks of this distribution are shifted toward higher angles. The X---O---X BAD is less affected by FS and also show a broad distribution centered around 80\textdegree\, for alkali silicate glasses and tend to increase with FS, while for alkaline earth silicate glasses the distribution is centered around 95\textdegree\, and show a secondary peak at around 110\textdegree\, only in the case of Mg and Zn silicate glasses. For the peaks between 40 and 70\textdegree\, results from the modifiers coordinated with two NBO or one NBO and one BO, which belongs to the same tetrahedron while the angle at 95\textdegree\, is attributed to the modifiers in octahedral geometry connecting two NBO from different tetrahedra \cite{Zhao2019}.

\begin{table}[t]
\caption{\label{Table:Rij}
Modifiers field strength and short-range structural parameters of glasses obtained from molecular dynamics at 300 K. The cutoffs was set to 2.1 and 2.9 \AA\, for the Si---O and O-O pairs, and 2.7, 3.3, 3.8, 2.67, 3.2, 3.5, 4.0, and 2.6 \AA\, for the Li---O, Na---O, K---O, Mg--O, Ca---O, Sr---O, Ba---O, and Zn---O pairs, respectively.}
\begin{tabular}{@{}llllllll}
\hline
Glass & FS & \multicolumn{2}{c}{$Si-O$}& \multicolumn{2}{c}{$O-O$} & \multicolumn{2}{c}{$X-O$}\\
\cline{3-8}
systems & (\AA\,$^{-2}$) &$r_{ij}$ & $N_{ij}$ & $r_{ij}$ & $N_{ij}$ & $r_{ij}$ & $N_{ij}$ \\
\hline
LiS & 1.49 & 1.614  & 4.00 & 2.61 & 5.30 & 1.96 & 3.80 \\
NaS & 0.69 & 1.614  & 4.00 & 2.61 & 5.20 & 2.32 & 5.98 \\
KS  & 0.39 & 1.614  & 4.00 & 2.62 & 5.19 & 2.69 & 7.56 \\
MgS & 3.12 & 1.602  & 4.00 & 2.62 & 5.42 & 2.01 & 4.38 \\
CaS & 1.51 & 1.602  & 4.00 & 2.61 & 5.27 & 2.33 & 6.03 \\
SrS & 1.13 & 1.614  & 4.00 & 2.61 & 5.25 & 2.55 & 7.12 \\
BaS & 0.89 & 1.614  & 4.01 & 2.60 & 5.27 & 2.77 & 10.68\\
ZnS & 3.12 & 1.602  & 4.00 & 2.62 & 5.30 & 1.96 & 3.86 \\
\hline
\end{tabular}
\end{table}

\begin{figure}[h!]
\centering
\includegraphics[width=\columnwidth]{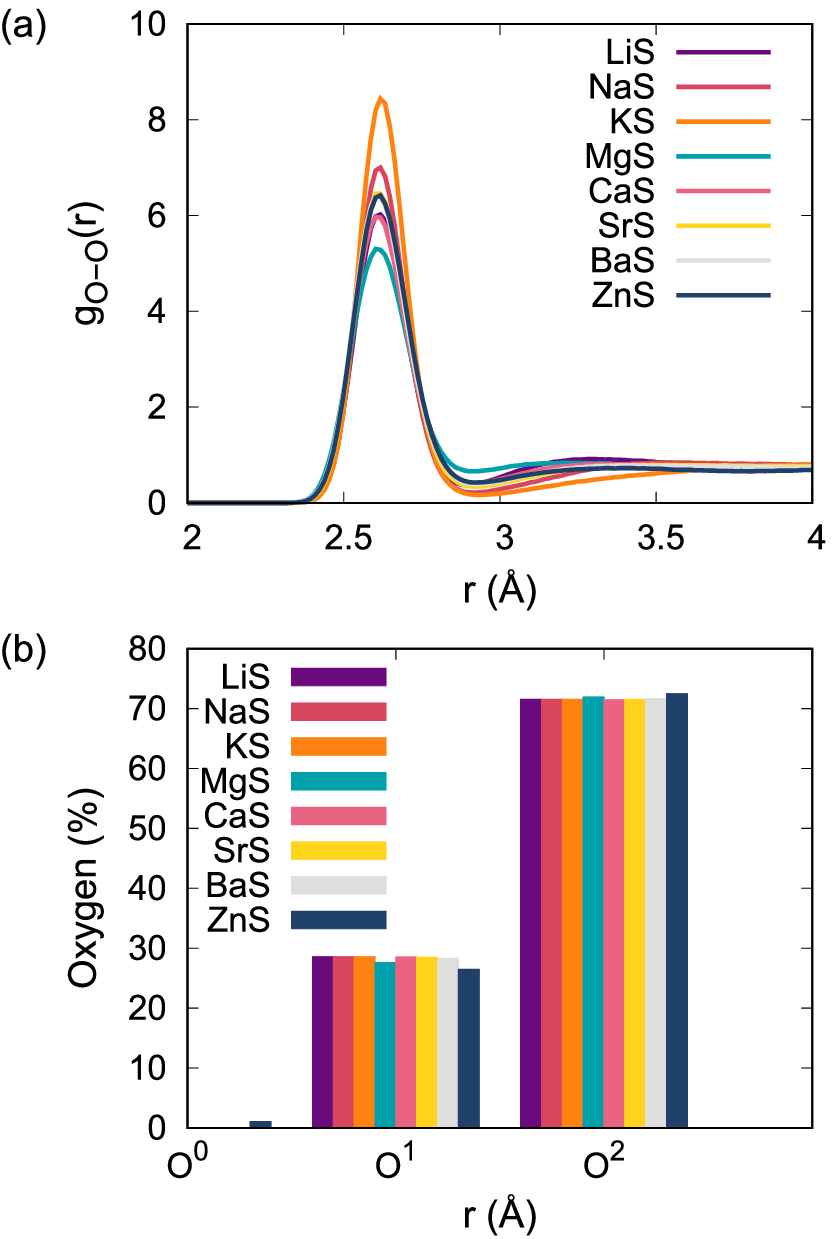}
\caption{(a) O-O radial distribution function, (b) oxygen species distribution in our glasses at 300 K.}
\label{fig:OO}
\end{figure}
The radial distribution function of the O---O pair is presented in Fig.~\ref{fig:OO}(a), which shows that the mean length of the O---O bond (which is given by the peak position) unaffected by the cations field strength (no shift of the RDF is observed) while the width of the first peak changes slightly with FS. This suggests the presence of longer O---O bonds in the systems with higher field strength. This can be seen as an indication of the presence of larger tetrahedra units in our systems with increasing FS.
\subsection{Medium-range structure}
In silica glass, the structure is made by tetrahedra connected by BO, which means that each Si is bonded to four O making a Q$^4$ structural unit and each O is bonded to two Si.  As stated previously, the addition of modifiers breaks the Si---O---Si bonds, and enabling the formation of NBO atoms. The identification of the oxygen species is made according to the total number of bonds to Si atoms. Thus, O$^0$, O$^1$, and O$^2$ stand for free oxygen (FO), non-bridging oxygen (NBO), and bridging oxygen (BO). As illustrated in Fig.~\ref{fig:OO}(b) the bridging oxygen have the highest population in the present silicate glasses; this is in agreement with previous experiments and molecular dynamics \cite{Atila2019a, Atila2019b, Ghardi2019, Matson1983, Huang1991}. Moreover, the existence of NBO in all glasses indicates that the network is depolymerized to some extent. No dependence on the field strength was seen in the evolution of the oxygen population. However, this may be due to the interatomic potential or the high cooling rate or even the sample size. 

\begin{figure}[htbt]
\centering
\includegraphics[width=\columnwidth]{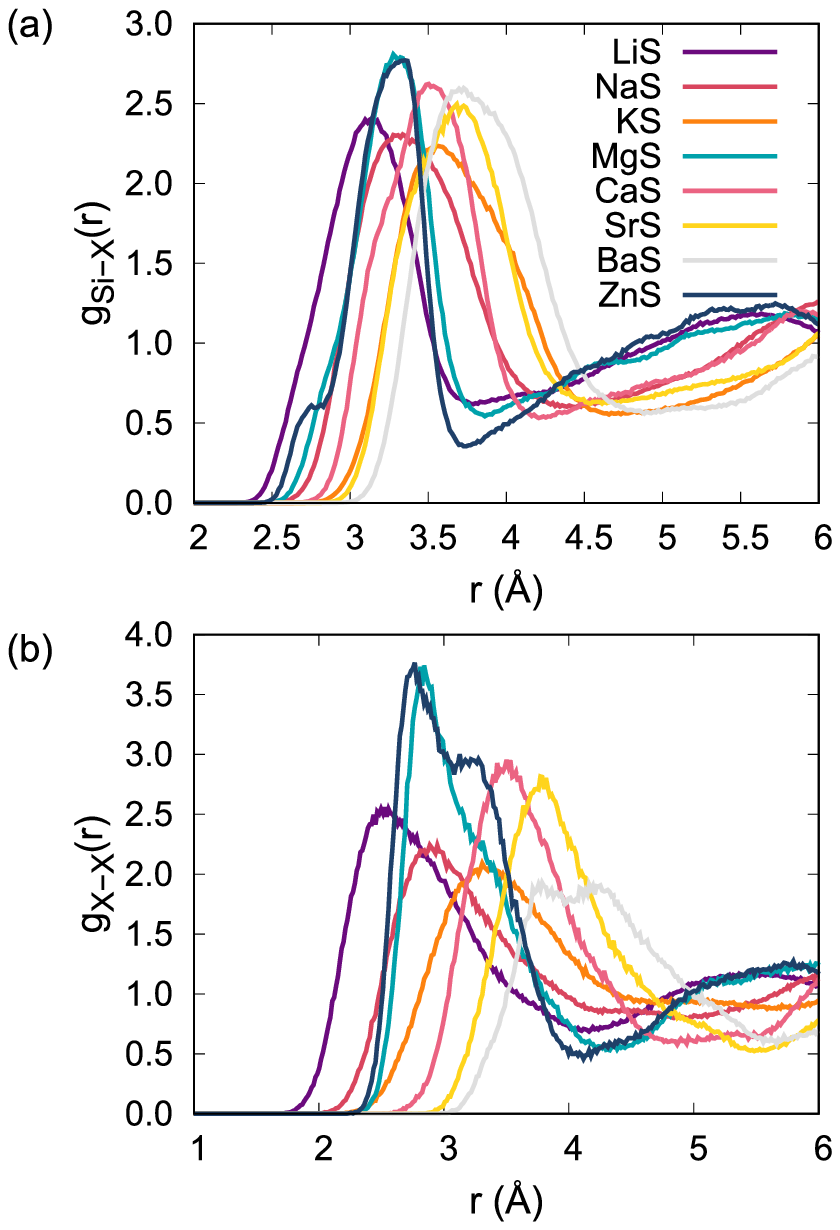}
\caption{(a) Si---X and (b) X---X radial distribution function silicate glasses as obtained from MD simulations at 300 K. X = Li, Na, K, Mg, Ca, Sr, Ba, and Zn. }
\label{fig:RDFSiX}
\end{figure}

To get more insights, we did compute and plot the RDF of the Si---X , as shown in Fig.~\ref{fig:RDFSiX}(a). We observe a negative correlation between FS and Si---X lengths.  This implies that modifiers with high FS tend to cluster around SiO$_4$ tetrahedral units. Additionally,  the X---X RDF show some peaks with decreasing first maximum as a function of FS as depicted in Fig.~\ref{fig:RDFSiX}(b).

\begin{figure}[htbt]
\centering
 \includegraphics[width=\columnwidth]{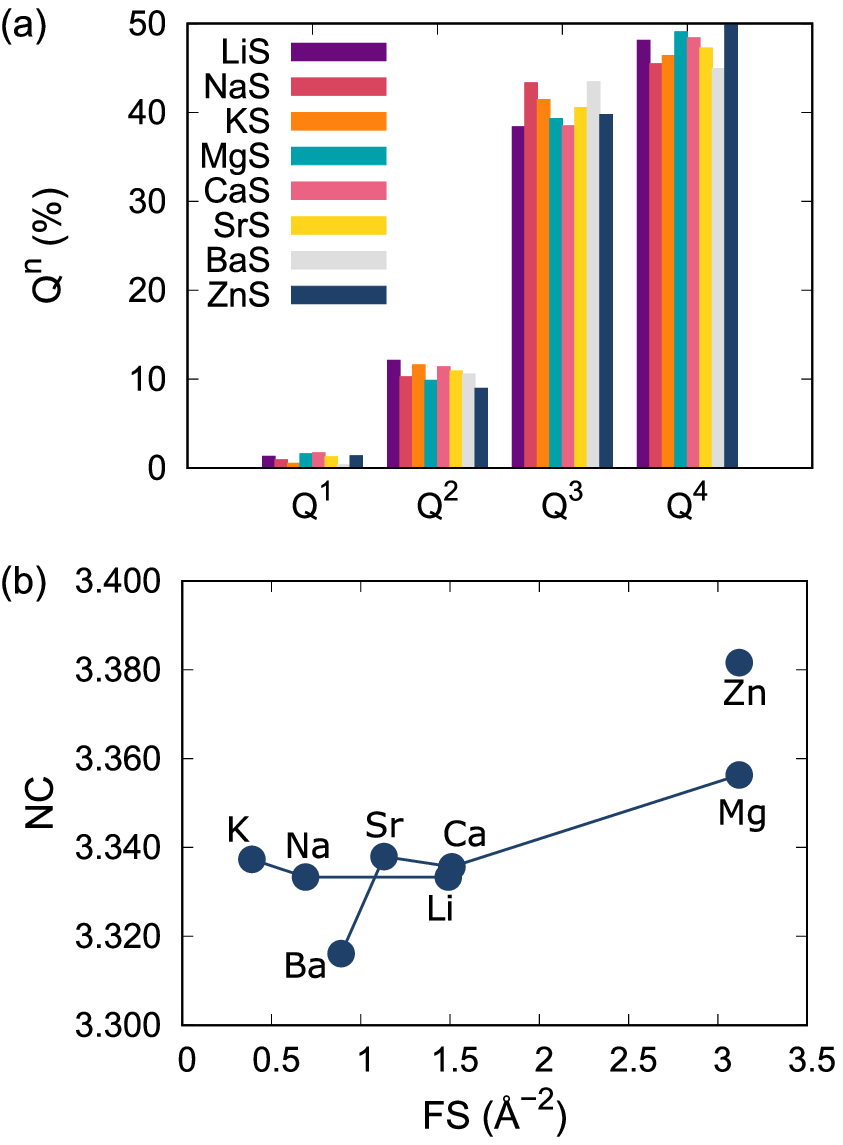}
 \caption{(a) Q$^n$ distribution and (b) network connectivity as calculated from the Q$^n$ distribution in the silicate glasses as obtained from MD simulations at 300 K.}
 \label{Qn}
\end{figure}
The medium-range structure of Si atoms can be further analyzed through the Q$^n$ distributions. In the modified silicate glasses, we have only one glass former, which is the silicon; thus, the Q$^n$ distribution is defined as the number of BO per Si tetrahedron. The Q$^n$ distribution is shown in Fig.~\ref{Qn}(a). The glasses show the co-existence of Q$^1$ to Q$^4$ species, with Q$^3$ and Q$^4$ representing the majority of the distribution.  The corresponding network connectivity (NC) is calculated using eq. \ref{meanQn}, and represents the average number of BO in the coordination shell of Si. Fig.~\ref{Qn}(b) depicts the network connectivity (NC) as a function of FS.
\begin{equation}
\label{meanQn}
NC = \sum_{n=1}^{4} x_nn
\end{equation}
where $x_n$ is the fraction of the Q$^n$ (n= 1, 2, 3, or 4) species. 
The NC tends to increase with increasing FS, indicating that the glass structure tends to be more polymerized when containing high FS modifiers.

\begin{figure}[htbt]
\centering
\includegraphics[width=\columnwidth]{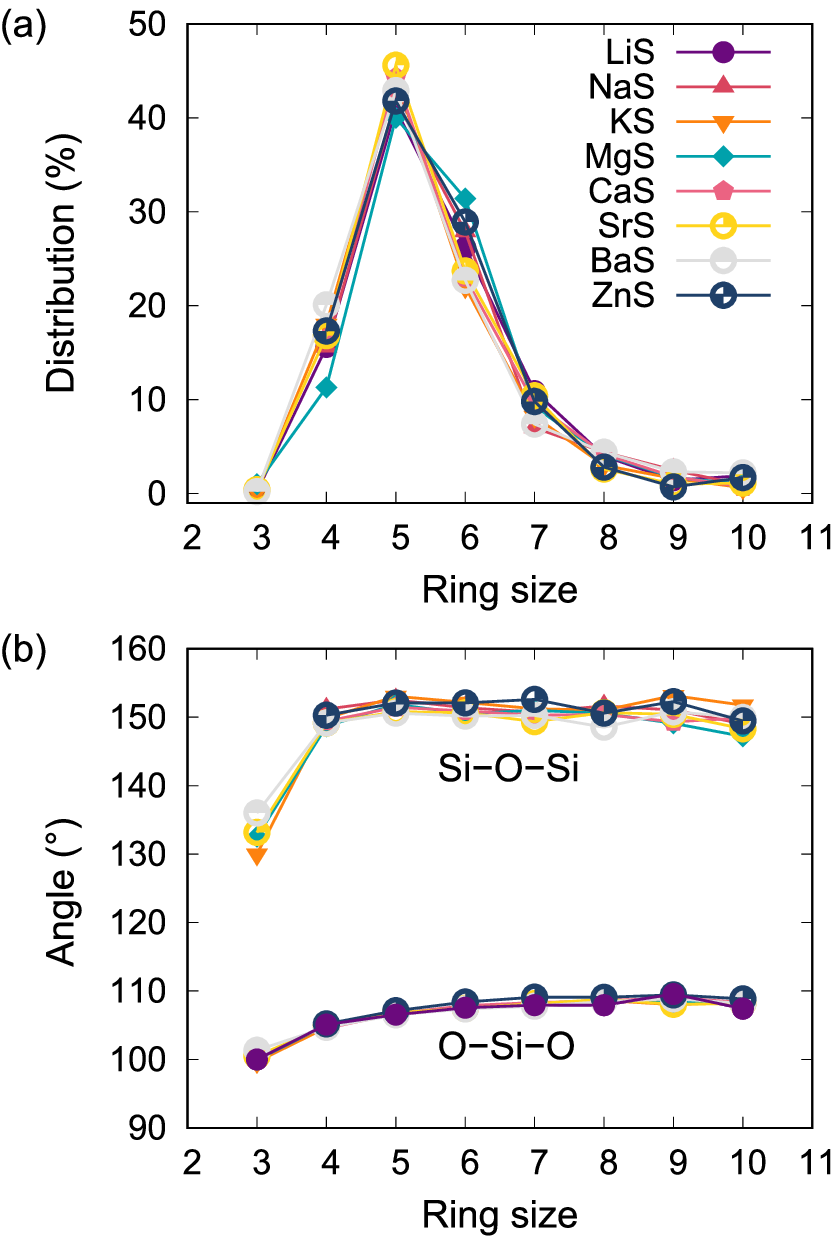}
\caption{(a) Ring size distributions per Si of the modified silicate glasses studied here. (b) Average Si–O–Si and O–Si–O angle in the modified silicate glass. The line is a guide to the eye.}
\label{Rings}
\end{figure}

Figure~\ref{Rings}(a) depicts the ring size distribution in the simulated silicate glasses. The ring size distribution was computed using the RINGS code \cite{LeRoux2010}. As can be seen, the ring size exhibits a broad distribution centered on the 5-membered rings in agreement with previous studies \cite{Atila2019a, Ghardi2019}. We did not observe any clear dependence of the ring size distribution on FS. 

In order to further analyze the structure through the ring distribution,  the average O---Si---O and Si---O---Si angles as a function of the ring size, as shown in Fig.~\ref{Rings}(b). We can see clearly that the average angles depend on the ring size. For small rings (3-membered rings), the O---Si---O and the Si---O---Si angles show values around 100 and 130\textdegree\,, respectively. Besides, these angles keep increasing up to the 6-membered rings where these angles show a constant value, which is equal to the mean O---Si---O and Si---O---Si calculated from the angular distribution function (See Fig.~\ref{SiOSi}(a and b)).  The small values of these angles in smaller rings were shown to indicate that small rings are strained and this is due to the over-constrained nature of these structures, leading to weaker angular bond bending constraints, which in turn yield to the stronger radial bond sharing constraints \cite{Song2019}.

\subsection{Modifiers clustering}
Now we focus our attention on how different modifiers are distributed in the glass matrix. The sharp peaks in radial distribution function of X---X (X = Li, Na, K, Mg, Ca, Sr, Ba, or Zn) (see Fig.~\ref{fig:RDFSiX}(b)) suggest some structuring, however, we cannot assume that this ordering is attributed to modifiers clustering. Hence, we studied the tendency of cluster formation (X---X clusters). Moreover, and to quantify and compare this tendency of modifiers clustering and relate it to the cation field strength, we compute the observed coordination number named after N$_{MD}$, which is defined as the integral of the RDF up to its first minimum, and compare it to the coordination number expected if the same density if ions were distributed randomly and uniformly throughout the glass N$_{hom}$ \cite{Christie2010, Christie2012, Tour2019, Tilocca2007}. This leads to the clustering ratio as defined by equation \ref{clusteringeq},

\begin{equation}
\label{clusteringeq}
R_{X-X} = \frac{N_{X-X,MD}}{N_{X-X,hom}} = \frac{CN_{X-X} + 1}{\frac{4}{3}\pi r_c^3 \frac{N_X}{V_{box}}}
\end{equation}

where R$_{X-X}$ is the clustering ratio, CN$_{X-X}$ is the modifier---modifier coordination number, N$_X$ stand for the total number of modifier atoms, r$_c$ the cutoff defined as the first minimum of the modifier-modifier RDF, and V$_{box}$ the volume of the simulation box. The ratio R$_{X-X}$ is now adopted to measure the degree of clustering. It is worth mentioning that the deviations from unity indicate clustering \cite{Christie2010}. In principle, the clustering ratio analysis could be done using any distance to check for longer-range clustering, but we used a distance up to the first coordination shell, to investigate the clustering in the first coordination shell. Another reason for using the first coordination shell is that in most cases, the second coordination shell is poorly defined and thus making the calculation of this ratio a non-trivial task \cite{Christie2010}.
\begin{figure}[htbt]
\centering
\includegraphics[width=\columnwidth]{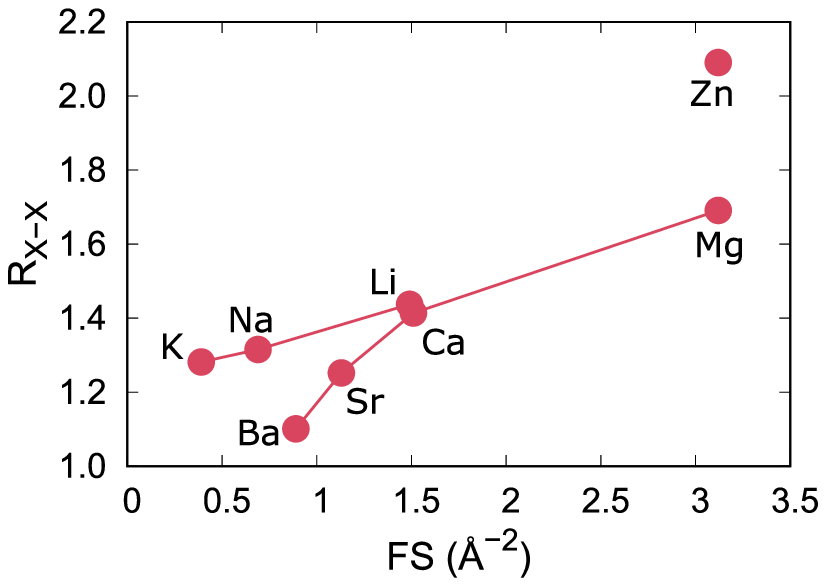}
\caption{The clustering ratio pf X---X pair as a function of the cations field strength. The lines are guide to the eye.}
\label{clusteringRatio}
\end{figure}
Fig.~\ref{clusteringRatio} shows the clustering ratios of the modified silicate glasses as a function of FS. All R$_{X-X}$ values are higher than unity, indicating that the clustering happens in all the simulated silicate glasses. Moreover, if we focus on the dependence of R$_{X-X}$ on FS, it is clear that they show almost a linear relationship, and this indicates that silicate glasses doped with cations having a higher field strength show a high tendency for modifiers clustering and thus phase separation. To support this, snapshots of the modifiers arrangements in the glassy matrix are shown in Fig.~\ref{clustering} were silicon and oxygen atoms were deleted for clarity. The figure shows clearly that the modifiers with high FS are clustered in some regions of the simulation box while the others with low FS tend to be more homogeneously distributed.

\begin{figure*}[htbt]
\centering
\includegraphics[width=\textwidth]{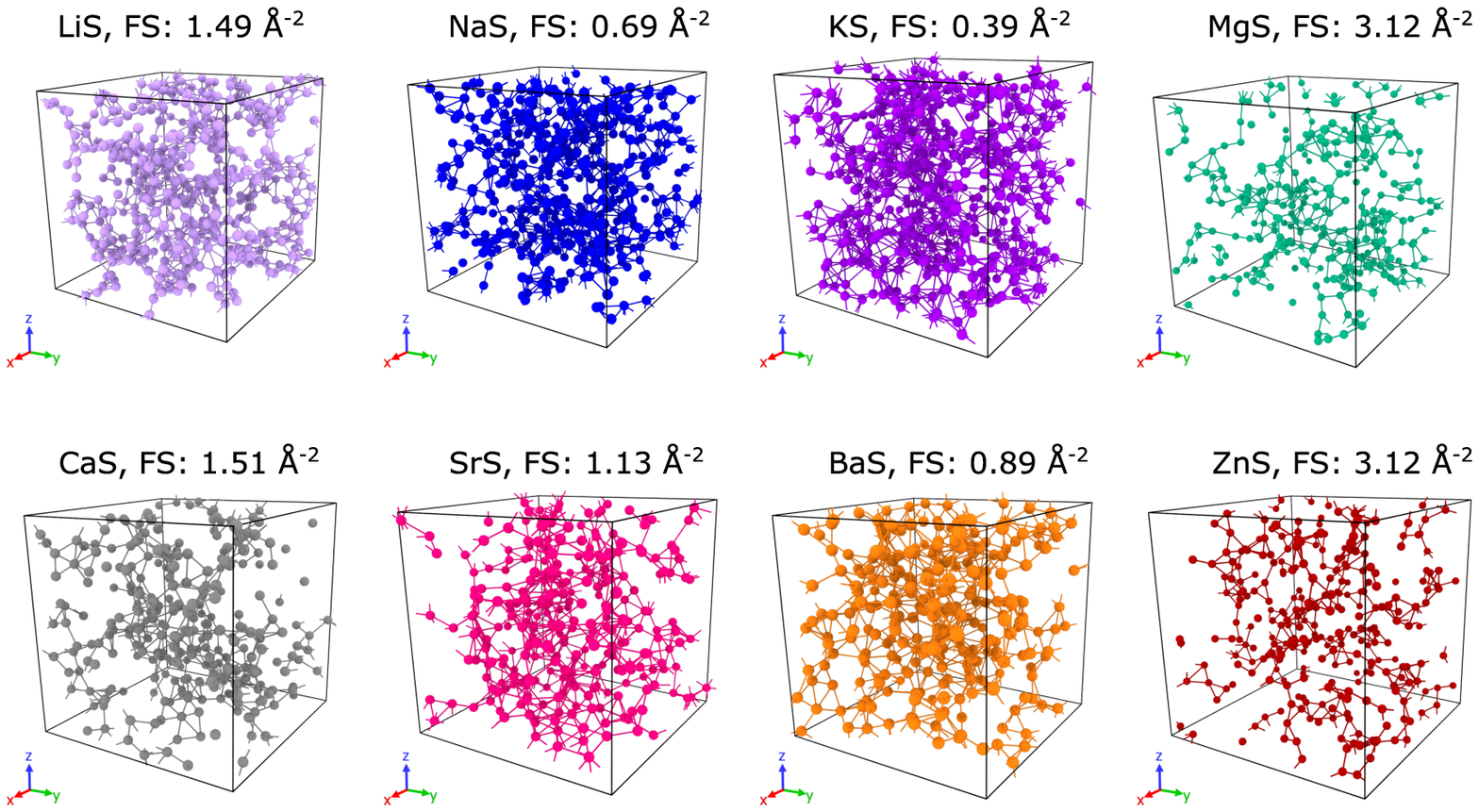}
\caption{Atomistic visualization of the modifiers atoms in  the studied silicates glasses. The clustering of the atoms is higher in the silicate glasses containing modifiers with a smaller size (e,g,. Li, Mg) (high FS) while a more random distribution of the modifiers is seen for the silicate glasses doped with larger cations (e,g,. K, Ba) (small FS). A 10 \AA\, slice is given in the bottom right of each sample. Only modifier atoms are shown for clarity.}
\label{clustering}
\end{figure*}

\section{\label{sec:Discussion}Discussion}
Overall, we did not observe any noticeable change in the mean bond lengths of Si---O; this is seen through the unchanged position of the RDF first peak. The modifier---oxygen bonds were shown to decrease with increasing FS. This is because smaller cations have less neighboring oxygen; thus, the bonds will be stronger and shorter. The effect of FS on X---O coordination numbers is clearly shown in Table. \ref{Table:Rij}, supporting the arguments saying that fewer oxygen atoms are present around high field strength cations. This implicates that each oxygen will have a more concentrated negative charge in order to balance the charge of the modifier. Moreover, this could be supported by the fact that this effect is more pronounced for divalent cations and not for monovalent cations. Consequently, higher field strength modifiers promote the concentration of negative charges on their local oxygen atoms. This, in fact, is in good accordance with the coordination number distribution and the size of the cation, as large cations have a larger surface area to distribute the charge, which can result in a high coordination number. Moreover, the O---X---O bond angle represents the angles inside XO$_n$ polyhedra, for the modifier showing a high FS value (Li, Mg, and Zn) the ADF showed a main peak around 100\textdegree\, and a secondary peak around 70\textdegree\,, the main peak is attributed to distorted tetrahedral units while the second peak is resulting from the modifiers coordinated with two NBO or one NBO and one BO which belongs to the same tetrahedron \cite{Zhao2019}. Moreover, for cations showing lower FS values, the BAD is broader and the main peaks are attributed to different polyhedral types (e,g,. the peaks at around 80\textdegree\, extending to 180\textdegree\, could be attributed to distorted octahedra \cite{Atila2019a}, and the secondary peaks at around 70\textdegree\, could be assigned to capped octahedra (CN$_{X-O}$ = 7). In addition, the X---X distances decrease with increasing FS, as expected. Also, We would like to stress that in our simulations, the short-range cation---cation repulsive forces are not considered except for Coulomb interaction. The differences between the cation---cation distances could be explained by the fact that the local environment of the cations is mostly defined by their oxygen coordination shells, and the short-range cation---cation repulsive effects are negligible if compared to the Coulombic repulsive effects \cite{Huang1991}. Thus, an indirect effect of the modifier---oxygen bond strength.

The analysis of the glasses through the Q$^n$ distribution and the network connectivity showed as expected that the highest population is that of Q$^3$ and Q$^4$ since the glasses should show a less polymerized network than in pure silica. Moreover, NC showed that the glasses containing modifiers with high FS tend to be slightly more polymerized, this could also be seen from the Q$^n$ distribution as for high field strength modifiers, and the tendency to form more Q$^4$ and Q$^1$ species than the cations with low FS, e,g,. lithium silicate glasses have Q$^2$ and Q$^4$ units than sodium and potassium silicate glasses, which evince the tendency of Li network modifier cations to cluster around Q$^2$ units in the glass. This will eventually lead to a local enrichment in Q$^4$ units and modifier clustering could be related to the presence of network modifier percolation channels in the glass structure \cite{Greaves1985, Greaves1995, Greaves2007}. Also, the rings analysis showed that the modified silicates glasses network contains rings made of SiO$_4$ tetrahedra linked by BO. As expected, the distribution of the rings is less affected by the type of modifier. Moreover, the distribution of the rings could also be explained by the presence of topological order/disorder. In general silica-rich glasses show a low fragility index, indicating their high glass-forming ability and this could be related to a higher topological disorder compared to poor silica glass. However, this could also be related to the type of the modifier present in the glass as the total number of rings tend to decrease with increasing FS (see supplementary materials Fig. 1 ), indicating the presence of the more disordered network. This could be due to the broad ring size distribution and the higher clustering ratio implicating increases the structural heterogeneity in the glass with FS. We could also get an estimate on the degree of ordering in the glass by computing the pair-excess entropy as defined in Ref. \cite{Atila2019b, Piaggi2017, Atila2020a}. As shown by the pair-excess entropy (S$_2$), the glass structure becomes more disordered as the FS increases (see Fig. 2 supplementary materials). We would like to stress that higher values of S$_2$ indicate higher configurational entropy and thus more disorder.

The analysis of the tendency of clustering showed that the peaks in the modifier---modifier RDF are indeed attributed to the modifiers forming clusters. Also, we emphasize that modifiers with a weak bonding to oxygen show a more homogeneous distribution and help in the homogenization of the glass. As the cation field strength increases (small cation size), the X---X coordination numbers decrease, and the modifier---modifier separation distance decreases, which will make the distribution of the modifiers in the glass network more clustered than the one assuming a homogeneous distribution. Higher deviations from unity were observed in the R$_{X-X}$ (e,g,. Zn and Mg); on the other hand, the barium silicate glass shows the lowest clustering ratio, which is near a uniform distribution and this is clearly seen in Fig.~\ref{clustering}. This is attributed to the nature of the interactions between modifiers (X---X). De Jong \textit{et al.} showed that the Li---Li interactions are attractive at a distance around the first minimum of Li---Li RDF, while at the same distance, the Na---Na interactions showed a neutral behavior and K---K interactions were repulsive \cite{deJong1981}. These results suggest that the alkali metals are not following the classical valence theory. Thus their interactions are not repulsive all the time. Indeed, we can link this behavior to the modifiers field strength, as can be understood from these results as the field strength increases the modifier---modifier interactions become more attractive which implicate that modifiers with high field strength choose to bond with a SiO$_4$ which already have one or more Si---O---X bonds. This will result in a separation of the network into two regions: a modifiers rich region and a SiO$_4$ in the form of Q$^4$ region supporting the fact that the clustering ratio increases with increasing FS. Using the same analogy, we can explain the behavior for the alkaline earth silicates glasses. The larger modifiers tend to be more homogeneously distributed in the glass network as larger modifiers prefer to bond with a SiO$_4$, which is not bonded to any other modifiers. Also, the degree of clustering will strongly depend on the composition of the glass, which is out of the scoop of the current paper.

%Indeed, the glass properties are affected by the cooling rate used in the glass preparation. However, We showed previously in the aluminosilicate glasses \citep{Atila2019b} that the bond lengths and  the coordination numbers of the glass former were unaffected by the cooling rates, which was in accordance with previous studies on the cooling rate effect in oxide glasses \cite{Tilocca2013, Li2017a, Atila2019b}. Finally, we would like to stress that the ability of modifier clustering could be strongly affected by the thermal and pressure history of the glass during the cooling process (cooling rate or pressure used in the glass preparation). In the present study we did not investigate this effect as in general the cooling rates used in molecular dynamics are higher than in experiments. It is is also worth noting that the glasses obtained by MD simulations are more random than the real glasses which is attributed to the high cooling rate used in MD \cite{Li2017a, Atila2019a, Atila2019b}, therefore, we expect that the effects observed and suggested from MD to be more pronounced in real glasses.

\section{\label{sec:Conclusion}Conclusion}
We used MD simulations to study the effect of the modifier field strength on the modifiers clustering and the structural properties of silicate glasses with the composition 0.25 (X$_{n/2}^{n+}$O)---0.75(SiO$_{2}$) where X is Li, Na, K, Mg, Ca, Sr, Ba, or Zn.\\
We have shown that the type of the modifier induces a structural change manifested by the change in the glass structure, as glasses, which contain modifiers with high field strength, tend to be more polymerized. Moreover, cations with similar FS behave similarly, as has been seen for MgO and ZnO silicate glasses. Also, we showed that the distribution of the rings is slightly affected by the type of modifiers. In what concerns the modifier clustering, we highlighted the effect of the type of modifier on the propensity for such clustering, and we showed that it increases with increasing FS. Low FS modifiers tend to be randomly distributed in the glassy matrix, thus effectively favoring the homogeneity of the glass network. Overall, our results suggest that the degree of local heterogeneity of the glass and the distributions of the network modifiers are correlated, and even if no phase separation is present, the microscopic structure of silicate glasses studied herein is not entirely random and shows some degree of ordering.\\
The insights obtained from the present molecular dynamics simulations to explain the effect of the modifier type on the structure and clustering ability will help in understanding the effect of FS on the glass structure and properties and in designing glass compositions for more advanced technological applications.

\section*{Acknowledgments}
I would like to thank Prof. A. Hasnaoui and Prof. S. Ouaskit for the fruitful discussions. The author gratefully acknowledge the computing resources provided by the Erlangen Regional Computing Center (RRZE) to run some of the simulations.

%\section*{Conflict of interest}
%The author declare that there is no known conflicts of interest.
\bibliography{biblio}
\end{document}